\documentclass[10pt]{iopart}
\usepackage[T1]{fontenc}
\usepackage[numbers,sort&compress]{natbib}
\usepackage{color}
\usepackage{amsmath}
\usepackage{amssymb}
\usepackage{graphicx}

\usepackage{parskip}

\begin{document}

\title{Strain-induced stacking transition in bilayer graphene}
\author{Nina C. Georgoulea$^{1}$, Stephen R. Power$^{1,2}$ and Nuala M. Caffrey$^{3, 4}$}
\address{$^1$ School of Physics, AMBER \& CRANN Institute, Trinity College Dublin, Dublin 2, Ireland}
\address{$^2$ School of Physical Sciences, Dublin City University, Dublin 9, Ireland}
\address{$^3$ School of Physics, University College Dublin, Dublin 4, Ireland}
\address{$^4$ Centre for Quantum Engineering, Science, and Technology, University College Dublin, Dublin 4, Ireland}
\ead{nuala.caffrey@ucd.ie}

\begin{abstract}
Strain, both naturally occurring and deliberately engineered, can have a considerable effect on the structural and electronic properties of 2D and layered materials. 
Uniaxial or biaxial heterostrain modifies the stacking arrangement of bilayer graphene (BLG) which subsequently influences the electronic structure of the bilayer. 
Here, we use Density Functional Theory (DFT) calculations to investigate the interplay between an external applied heterostrain and the resulting stacking in BLG. 
We determine how a strain applied to one layer is transferred to a second, `free' layer and at what critical strain the ground-state AB-stacking is disrupted. 
To overcome limitations introduced by periodic boundary conditions, we consider an approximate system consisting of an infinite graphene sheet and an armchair graphene nanoribbon (AGNR). 
We find that above a critical strain of $\sim 1\%$, it is energetically favourable for the free layer to be unstrained, indicating a transition between uniform AB-stacking and non-uniform mixed stacking. 
This is in agreement with a simple model estimate based on the individual energy contributions of strain and stacking effects. 
Our findings suggest that small levels of strain provide a platform to reversibly engineer stacking order and Moir{\'e} features in bilayers, providing a viable alternative to twistronics to engineer topological and exotic physical phenomena in such systems.
\end{abstract}

\ioptwocol
\section{Introduction}
Two-dimensional (2D) materials, such as graphene, exhibit unique mechanical and electronic properties~\cite{akinwande2017review, kim2020heterogeneous, novoselov2005two, katsnelson2007graphene, lee2008measurement}.
2D materials can further be combined to create heterostructures which can have different properties to their component layers due to interlayer interactions~\cite{androulidakis2018tailoring, wang2014two}.
For example, the 2D elastic moduli of bilayer heterostructures, such as graphene/MoS$\mathrm{_{2}}$ and MoS$\mathrm{_{2}}$/WS$\mathrm{_{2}}$, are smaller than the sum of the moduli of the individual layers~\cite{liu2014elastic}. Interlayer interactions can be tuned, for example, by changing the relative stacking of the layers.
This allows a wide range of different behaviours to be observed, even in structures with multiple layers of the same material.

\begin{figure}[t]
\centering
\includegraphics[width = 1.0\linewidth, keepaspectratio]{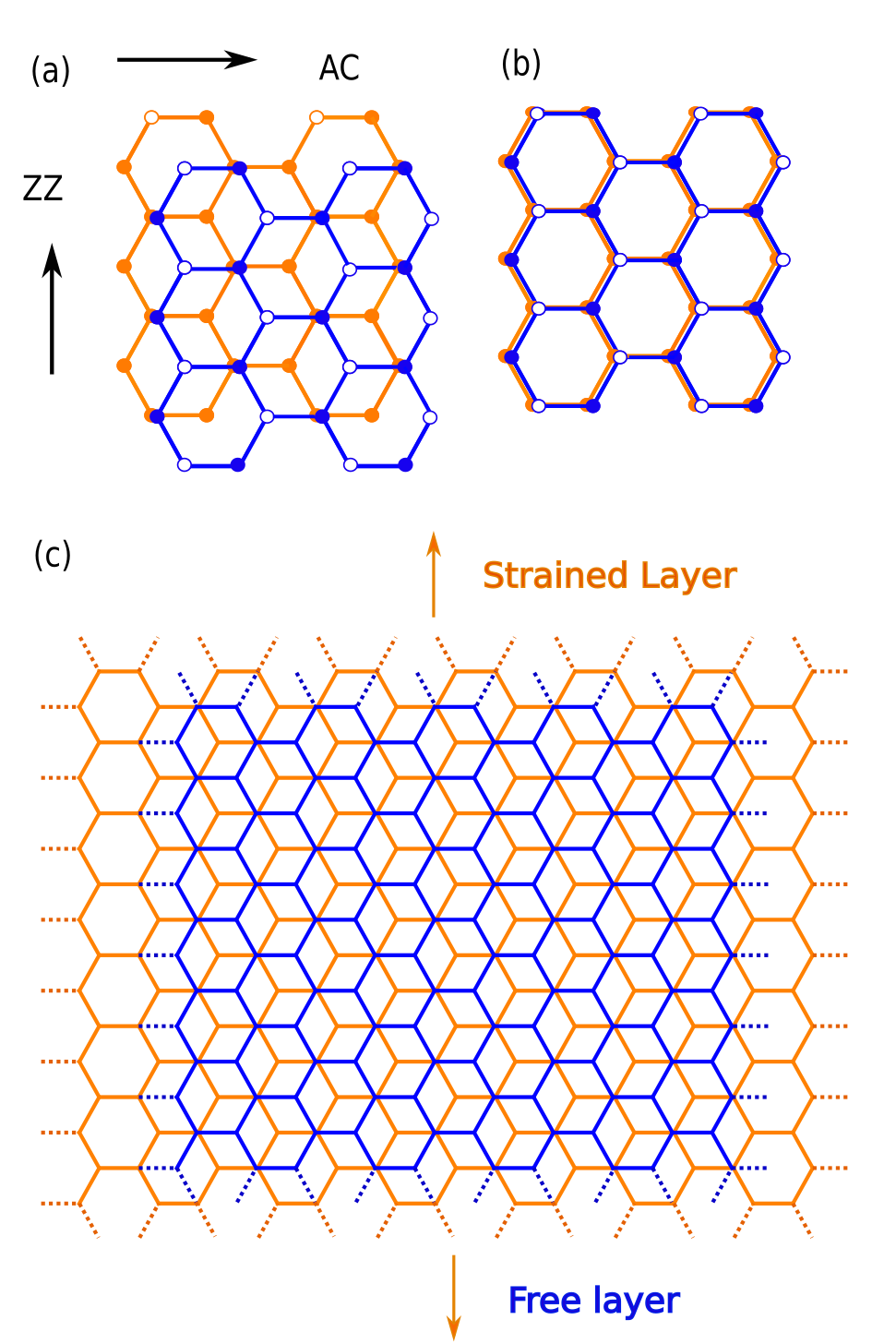}
\caption{(a) Structure of AB-stacked BLG, showing the high symmetry armchair (AC) and ZZ directions. The filled and hollow symbols represent atoms from each of the two sublattices. (b) Structure of AA-stacked BLG. (c) Structure of the BLG system considered in this work, where uniaxial strain is applied only to the bottom layer (orange) and we consider strained and unstrained configurations of the free layer (blue).}
\label{fig:general}
\end{figure}

Two monolayer graphene layers (MLGs) can be stacked to form bilayer graphene (BLG)~\cite{novoselov2004electric, ohta2006controlling}.
In the ground-state AB-stacking, half of the carbon atoms in each layer are directly above the centre of a hexagon on the other layer and the other half are directly on top of another carbon atom~\cite{mccann2013electronic, rozhkov2016electronic} (figure \ref{fig:general}(a)). 
AA-stacking has also been observed in which every carbon atom of one layer is directly above a carbon atom on the second layer (figure \ref{fig:general}(b)).
AB-stacked BLG has parabolic electronic bands, whereas the bands of AA-stacked BLG remain linear, as in MLG~\cite{ho2006coulomb, novoselov2004electric}. As a result, AB- and AA-stacked BLG behave very differently under the application of interlayer bias, which opens a band gap proportional to the bias for AB-stacking~\cite{castro2007biased}, while the AA-system remains semimetallic~\cite{silva2020electronic}.
Between the AB- and AA- stacking limits lie a range of different stacking possibilities. The electronic and topological properties of these systems varies as a function of relative shift between the layers~\cite{bhattacharyya2016lifshitz, son2011electronic}.
For example, the transport properties of BLG depend sensitively on its stacking, with a large change in the transmission predicted when one layer is shifted relative to the other~\cite{bhattacharyya2016lifshitz}. 
Thus, manipulating the stacking of a bilayer is a powerful tool to tune its electronic and transport behaviour.
Twisted BLG (TBLG) harnesses this to great effect~\cite{dos2007graphene, shallcross2010electronic, dai2016twisted, mcgilly2020visualization}.

Both uniform and non-uniform strain can tune the electronic, transport and optical properties of both MLGs and BLGs~\cite{ni2008uniaxial, pereira2009tight, ni2010anisotropic, pellegrino2010strain, mucha2011strained, gradinar2012conductance, de2012space, zhang2014transport, Settnes2016}.
Strain can arise naturally when graphene is placed on a substrate due to lattice mismatch between the two systems~\cite{aitken2010effects, yang2021strain}. 
Strain can also be intentionally created and controlled in graphene by using a flexible substrate~\cite{yu2008raman, huang2009phonon}.
If BLG is placed on a flexible substrate, strain will be applied to the layer which is in direct contact with the substrate~\cite{androulidakis2020tunable}.
It is reasonable to expect that if the applied strain is sufficiently small, it will be entirely transferred to the second graphene layer, i.e., both layers will experience the same strain. 
However, for larger applied strains the second layer can exhibit a different strain profile~\cite{frank2012phonon, wang2019robust, androulidakis2020tunable}.
A heterostrain modifies the stacking between layers in a manner similar to a relative twist angle, creating a Moir{\'e} superlattice~\cite{choi2010controlling, van2016piezoelectricity, parhizkar2022strained}. 
Theoretical works predict that heterostrain can be used to open and tune an electronic energy gap in BLG~\cite{choi2010controlling}, or to induce a transition from a direct to an indirect band gap in the presence of a bias field~\cite{crosse2014strain}.

In this work, we investigate the interplay between uniaxial heterostrain and stacking effects in BLG.
The system consists of a bottom layer which is uniaxially strained along the zigzag (ZZ) direction and a `free' top layer as shown in figure \ref{fig:general}(c). 
For small amounts of strain applied to the bottom layer, the free layer is expected to strain by the same amount, as the energetic cost of straining is small compared to that of breaking uniform AB-stacking. 
However, at a certain critical value of applied strain, the energetic cost of maintaining strain in the free layer exceeds the cost of breaking AB-stacking. 
At this point, strain is released in the free layer as the energy benefit in doing so is greater than the energy penalty to be paid for a less-than-ideal stacking configuration. 
At this critical strain, a transition between a uniform AB-stacking order and a non-uniform stacking will occur, together with the formation of Moir{\'e} superlattices, as in TBLG~\cite{bistritzer2011moire, dai2016twisted}.
Near this critical point, the change in stacking order can have significant implications on electronic properties~\cite{bhattacharyya2016lifshitz, son2011electronic, choi2010controlling}.

\section{Computational Details}\label{sec:2}
Density functional theory calculations were performed using \textsc{vasp}-5.4.1~\cite{kresse1993ab, kresse1994ab, kresse1996efficiency, kresse1996efficient, blochl1994projector, kresse1999ultrasoft}. The Perdew-Burke-Ernzerhof (PBE)~\cite{perdew1996generalized} parametrization of the generalized gradient approximation (GGA) was employed. Van der Waals (vdW) interactions are included using the D2 semi-empirical method of Grimme~\cite{grimme2006semiempirical}. The plane wave basis set was converged using an 950~eV energy cutoff. A 13$\times$21$\times$1 mesh was used to determine the total energies of MLG and BLG. All structures were optimized until the residual forces were less than 0.01 eV/\AA.
A vacuum layer of at least 11.5~\AA\ was included in the direction normal to MLG or BLG to ensure no spurious interactions between repeating slabs.
The GGA calculated lattice constants of MLG and BLG were both found to be 2.47~\AA, in good agreement with the experimental value of 2.46~\AA.
The interlayer distances in AB- and AA-stacked BLG were found to be 3.37~\AA\ and 3.50~\AA, respectively, in good agreement with previous studies~\cite{tao2012comparative, lee2008growth, dahn1990suppression}.

To overcome restrictions caused by the use of periodic boundary conditions in the graphene plane, the heterostrained BLG system was approximated using a hydrogen-terminated armchair graphene nanoribbon (AGNR) adsorbed on top of a MLG. The dangling C bonds at the AGNR edges were passivated with H atoms. Heterostrain was then introduced by straining the MLG along the ZZ direction. A 21$\times$5$\times$1 k-point mesh was used to converge the total energy of the heterostrained structures.
A distance of at least 12.5~\AA\ was maintained between periodic replicas of adsorbed AGNRs to ensure that they do not interact.
The optimal AGNR width was determined by comparing the stacking-dependent AGNR binding energy ($E_B$) to the stacking-dependent binding energy of BLG. In this case, the in-plane positions of the two central atoms of the AGNR were held fixed at the chosen stacking and all other AGNR atoms were allowed to relax both in-plane and out-of-plane.
To determine the lowest energy stacking configuration for different heterostrains, the carbon-hydrogen bonds and the interlayer distance between the MLG and the AGNR were relaxed while the in-plane positions of all other atoms were held fixed.

\section{Results}\label{sec:3}

To determine the critical strain for which a transition occurs from a uniformly strained BLG with AB-stacking to a heterostrained system with disrupted stacking, we compare the energetics of two limiting cases: when the free layer either adopts the same strain as that applied to the bottom layer, or it remains completely unstrained. 
Possible intermediate scenarios, where the free layer adopts a non-zero strain different to that in the bottom layer or displays a non-uniform strain distribution, are not considered in this work due to their high computational cost.
Similarly, in order to maintain periodicity for the DFT calculations, we neglect the role of contraction in the direction perpendicular to the applied strain.
This is equivalent to setting the Poisson ratio $\nu=0$.
We discuss how these approximations can be relaxed, and the expected consequences, in Section \ref{sec:4}.

\subsection{Simple Model} \label{sub:1}
There are two principal energy costs, due to strain ($\Delta E_{\mathrm{strain}}$) and stacking ($\Delta E_{\mathrm{stack}}$), which determine the behaviour of the free layer when uniaxial strain is applied to the bottom layer.
To get a rough estimate of where the transition between a strained and unstrained free layer occurs, we can compare the expected energy costs of straining the free layer in isolation, and of breaking AB-stacking in an unstrained BLG system.

\begin{figure}
\centering
\includegraphics[width=1.0\linewidth, keepaspectratio]{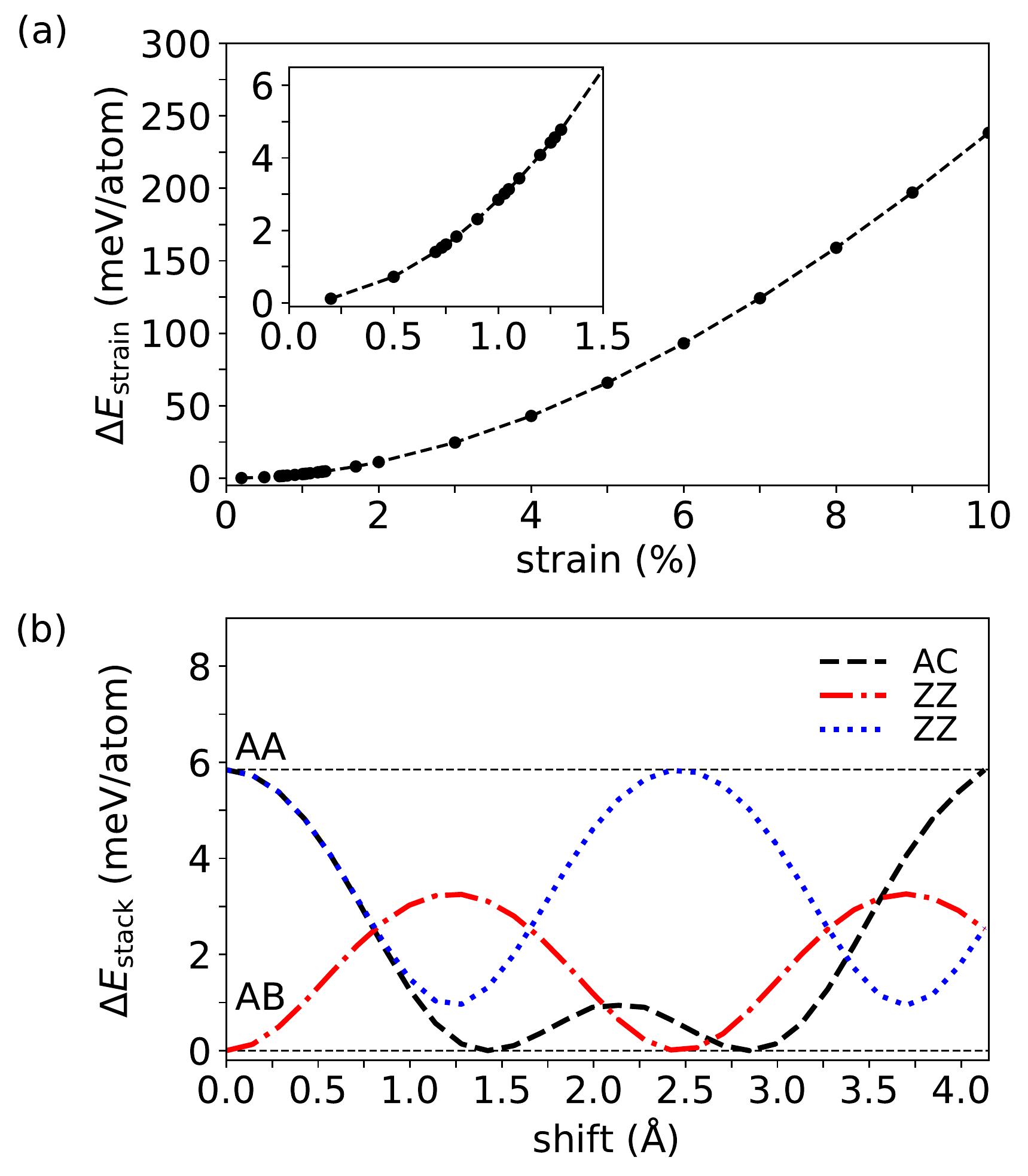}
\caption{(a) $\Delta E_{\mathrm{strain}}$ per atom as a function of applied ZZ uniaxial strain in MLG. (b) $\Delta E_{\mathrm{stack}}$ per atom as a function of the relative shift between layers in BLG for different initial stackings (AA and AB-stackings). Shifts along both ZZ and AC directions are shown.}
\label{fig:both}
\end{figure}

We first consider $\Delta E_{\mathrm{strain}} = E - E_{\mathrm{unstrained}}$, the energy cost associated with straining the free layer away from its relaxed structure to match the strain applied to the bottom layer.
The energy cost of straining a graphene layer increases with the amount of strain considered, as shown in figure \ref{fig:both}(a) for MLG with strain applied along the ZZ direction.
$\Delta E_{\mathrm{strain}}$ displays almost identical behaviour in AA- and AB-stacked BLG, and for strains along the AC direction in all three systems, with a maximum variation of only 2.8 meV per atom.
Since $\Delta E_{\mathrm{strain}}$ is not significantly affected by the strain direction, or the nature of the stacking in BLG systems, the curve in figure \ref{fig:both}(a) should also be an excellent approximation to the energetic cost of straining a single layer in BLG in the absence of stacking effects.

If the amount of strain is different in the two layers, then the system is no longer able to maintain energetically-favourable AB-stacking, and instead must display a modulation of the stacking order with an associated energy cost $\Delta E_{\mathrm{stack}} = E - E_\mathrm{AB}$.
Although the modulation wavelength depends on the strain mismatch, the energy cost per atom is roughly constant, as a similar range of stackings will occur for any mismatch.
Therefore, $\Delta E_{\mathrm{stack}}$ should not depend sensitively on the strain applied to the bottom layer, and we approximate it by considering different stacking configurations in unstrained BLG. 
Figure \ref{fig:both}(b) shows the energetic cost of rigidly shifting one graphene layer over the other along ZZ or AC directions, starting from either an initial AB- or AA-stacking. 
As the layers are shifted, the in-plane positions of atoms are held fixed to maintain the desired stacking configuration, but the interlayer distance is allowed to relax.
Any shift away from AB-stacking results in a positive $\Delta E_{\mathrm{stack}}$, confirming that this is the preferred configuration.
$\Delta E_{\mathrm{stack}}$ is maximum for AA-stacking with a value of 5.84 meV/atom, in excellent agreement with Ref. \cite{bhattacharyya2016lifshitz}.
The stacking modulation arising from heterostrain contains a combination of various stackings, and the corresponding $\Delta E_{\mathrm{stack}}$ can be approximated as an appropriately weighted average of the values appearing in figure \ref{fig:both}(b).

Comparing figure \ref{fig:both}(a) and \ref{fig:both}(b) allows us to understand the interplay between strain and stacking in heterostrained BLG. 
For small strains applied to the bottom layer, $\Delta E_{\mathrm{strain}} \ll \Delta E_{\mathrm{stack}}$, and it is energetically favourable for the free layer adopt the same strain.
However, as the strain applied to the bottom layer is increased, the cost of uniformly straining the free layer eventually balances the cost of breaking AB-stacking. 
A transition occurs above this critical strain, releasing the strain in the free layer and introducing a modulation of the stacking order. 
The maximum possible value of critical strain is restricted by the finite range of $\Delta E_{\mathrm{stack}}$: the maximum possible cost of breaking AB-stacking (i.e., $\Delta E_{\mathrm{stack}}^{AA} \sim 5.84$ meV/atom), corresponds to a strain of 1.4\%. 
In reality, the mix of different stackings that occur in a hetero-strained BLG will give $0 < \Delta E_{\mathrm{stack}} < \Delta E_{\mathrm{stack}}^{AA}$. 
For an even distribution of stackings between AB and AA, we can estimate $\Delta E_{\mathrm{stack}} \sim \frac{1}{2} \Delta E_{\mathrm{stack}}^{AA}$, corresponding to a critical strain of 0.97\%.
Uneven stacking distributions can occur if the considered strain excludes certain stackings, or if a non-uniform strain distribution is allowed in the free layer.
However, even accounting for a significant reduction in $\Delta E_{\mathrm{stack}}$ due to these effects does not dramatically change the expected critical strain. 
For example, assuming $\Delta E_{\mathrm{stack}} \sim \frac{1}{4} \Delta E_{\mathrm{stack}}^{AA}$ still gives a critical strain of 0.66\%.
The results of this simple model strongly suggest that the critical strain is near $1\%$, and that uniform AB-stacking will be broken when larger strains are applied to the bottom layer.

\subsection{AGNR on strained MLG}
DFT calculations of heterostrained bilayers will now be used to test the prediction of this simple model that the critical strain occurs near 1\%.
However, periodic boundary conditions enforce a commensurability condition when dealing with two infinite graphene sheets. 
Although neglecting contraction in the transverse direction ($\nu=0$) simplifies matters considerably, only certain values of strain would be achievable~\cite{choi2010controlling}.
Furthermore, very large supercells would be required to investigate the relevant strain range: $1\%$ strain requires 101 cells of the free layer and 100 cells of the strained layer.
Getting sufficient resolution to determine the critical strain quickly becomes computationally prohibitive. 

\begin{figure}[t]
\centering
\includegraphics[width=1.0\linewidth,  keepaspectratio]{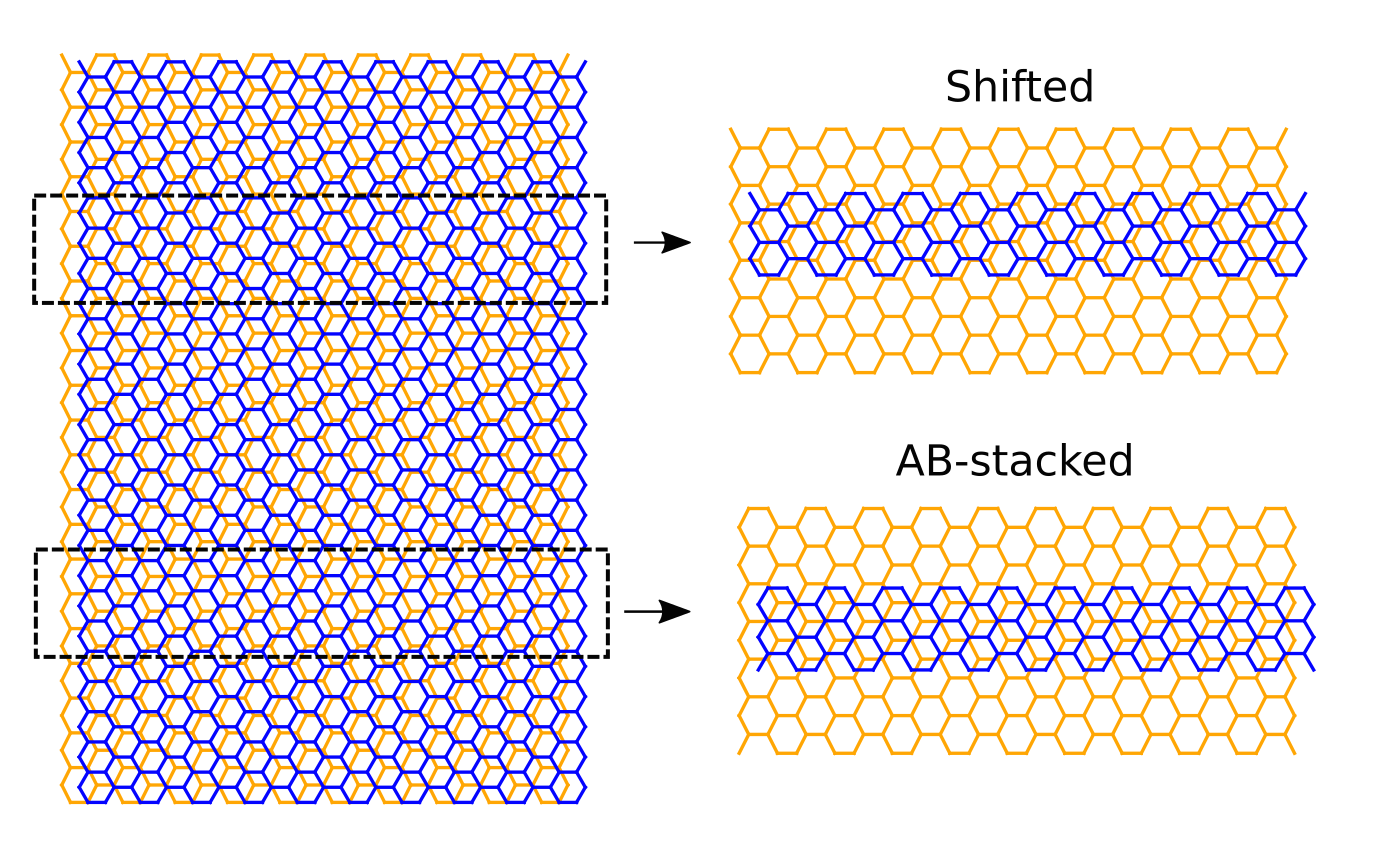}
\caption{ Left: Structure of BLG when the bottom (orange) layer is uniaxially strained by 15\% along the ZZ (vertical) direction. A modulation of the stacking is clearly visible. Right: The systems considered in this work, where AB and shifted regions of the BLG system are represented by finite-width AGNRs.}
\label{fig:blg}
\end{figure}

To overcome this constraint, we instead investigate the interplay of strain and stacking in a system where different regions of the free layer are modelled by finite-width AGNRs.
This is shown schematically on the left side of figure \ref{fig:blg}, for an exaggerated strain of 15\% along the ZZ (vertical) direction applied to the bottom layer.
A one-dimensional Moir{\'e} pattern is evident in the full BLG with regions of AB-stacking (lower dashed box) separated by other stacking types. 
Due to the ZZ strain direction and $\nu=0$ , no AA-stacking occurs and the furthest stacking from AB is that in the upper dashed box, which we denote `Shifted' and corresponds to rigidly shifting one layer of unstrained BLG by half a graphene lattice constant in the ZZ direction.
We aim to determine the energetics of the complete system (left hand side) by modelling different portions of it by a finite-width ribbon adsorbed onto an infinite bottom layer (right hand side).
The AGNR can be rigidly shifted over the continuous bottom layer to approximate the different stackings that occur in a heterostrained bilayer system.
As the free layer is no longer continuous, but now consists of a periodic array of AGNRs, we can consider different strains in each layer using a constant-size supercell.
We consider hydrogen-passivated AGNRs to circumvent features including unpassivated bonds or localised edge states. These may occur in certain AGNRs but are not expected in extended bilayer systems~\cite{raza2011edge, nakada1996edge, fujita1996peculiar}.

\begin{figure}
\includegraphics[width=1.0\linewidth, keepaspectratio]{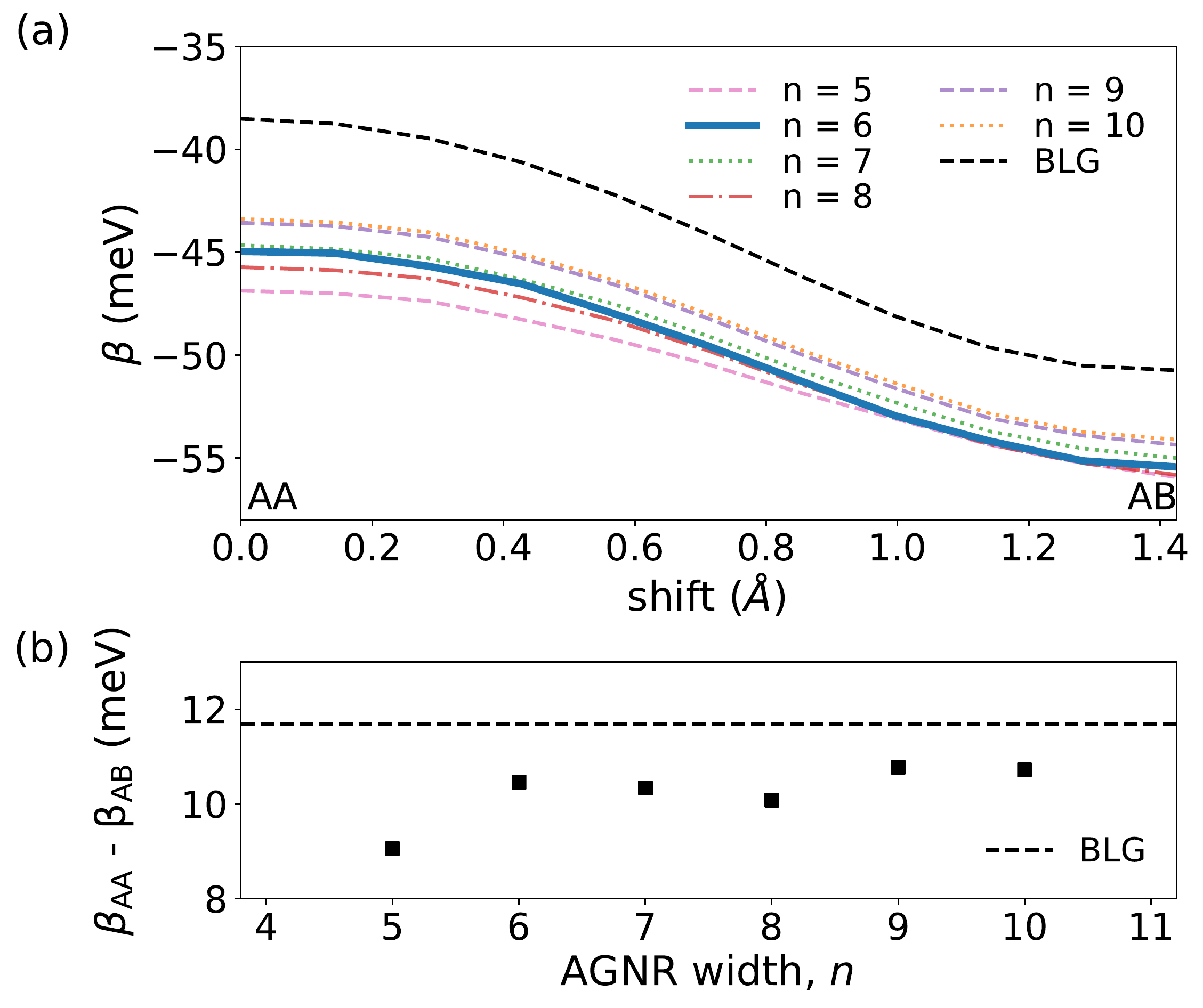}
\caption{(a) $\beta$ (i.e., $E_B$ per overlap atom) as a function of stacking for different width $n$-AGNR/MLG systems. (b) Energy difference $\beta_{AA}-\beta_{AB}$ as a function of width $n$ for \textit{n}-AGNR/MLG. The black dashed lines show the corresponding infinite BLG results.}
\label{fig:first1}
\end{figure}

The choice of AGNR width is determined by that which best approximates the stacking-dependent 
binding energy of BLG at a reasonable computational cost.
To meaningfully compare the binding energy of BLG and the $n$-AGNR/MLG systems we normalise $E_B$ by the number of carbon atoms in the top-layer, $N_C$:
\begin{equation}
\beta = \frac{E_B}{N_C}
\end{equation}
Figure \ref{fig:first1}(a) shows how $\beta$ varies as the top layer is rigidly shifted along the AC direction from AA to AB alignment with the bottom layer. 
The dashed black curve shows $\beta$ for BLG while the other curves show $\beta$ for the $n$-AGNR/MLG systems with $n=5,\dots, 10$.
While there is an offset between the BLG binding energy and that of the AGNR/MLG, this offset is approximately constant across all stackings between AA and AB. For widths in excess of $n=6$ the error is less than 10\%. 
Figure \ref{fig:first1}(b) shows the binding energy difference between AA and AB-stackings, $\beta_{AA}-\beta_{AB}$, as a function of width $n$, compared to the corresponding quantity for BLG (dashed line). 
Agreement between the BLG and AGNR/MLG systems improves in general with the increase of $n$, but non-uniformly due to different behaviour of ribbons with widths $n=3q, 3q+1$ and $3q+2$, where $q=1, 2, 3, \dots$.  
Similar trends have been noted, for example, for the band gap of AGNRs~\cite{wakabayashi1999electronic}. 
As we are interested in how stacking changes the energetics, and not the absolute magnitude of the $E_B$, the $6$-AGNR is deemed sufficiently wide for our purposes. 

To estimate the critical strain in heterostrained BLG using the 6-AGNR/MLG system, we consider the energy difference, $\Delta E_X$, between strained and unstrained AGNR layers, for different stackings ($X=$ AB, shifted).
We emphasise that it is the \emph{unstrained} case which gives rise to broken stacking, due to the strain applied to the bottom layer, whereas the \emph{strained} case restores AB-stacking by matching the strain in the both layers.
The energy difference between strained and unstrained AGNR layers is then given by:
\begin{equation}
  \Delta E = E_{\mathrm{AB,\,strain}} - E_{X,\mathrm{unstrain}}  \,,
  \label{eqn:deltaE}
\end{equation}
where we note that the energy of the final state $E_{\mathrm{AB,\,strain}}$ is the same in each case, as AB-stacking has been restored.
$E_{X, \mathrm{unstrain}}$ corresponds to the case that the MLG is strained and the AGNR is unstrained, with the stacking, $X$, set by fixing the positions of the central carbon atoms of the AGNR. 
The in-plane positions of the AGNR carbon atoms are held fixed as determined by the strain and stacking, while the hydrogen-carbon bonds and the interlayer distance are allowed to relax.

\begin{figure}
\centering
\includegraphics[width=1.0\linewidth,  keepaspectratio]{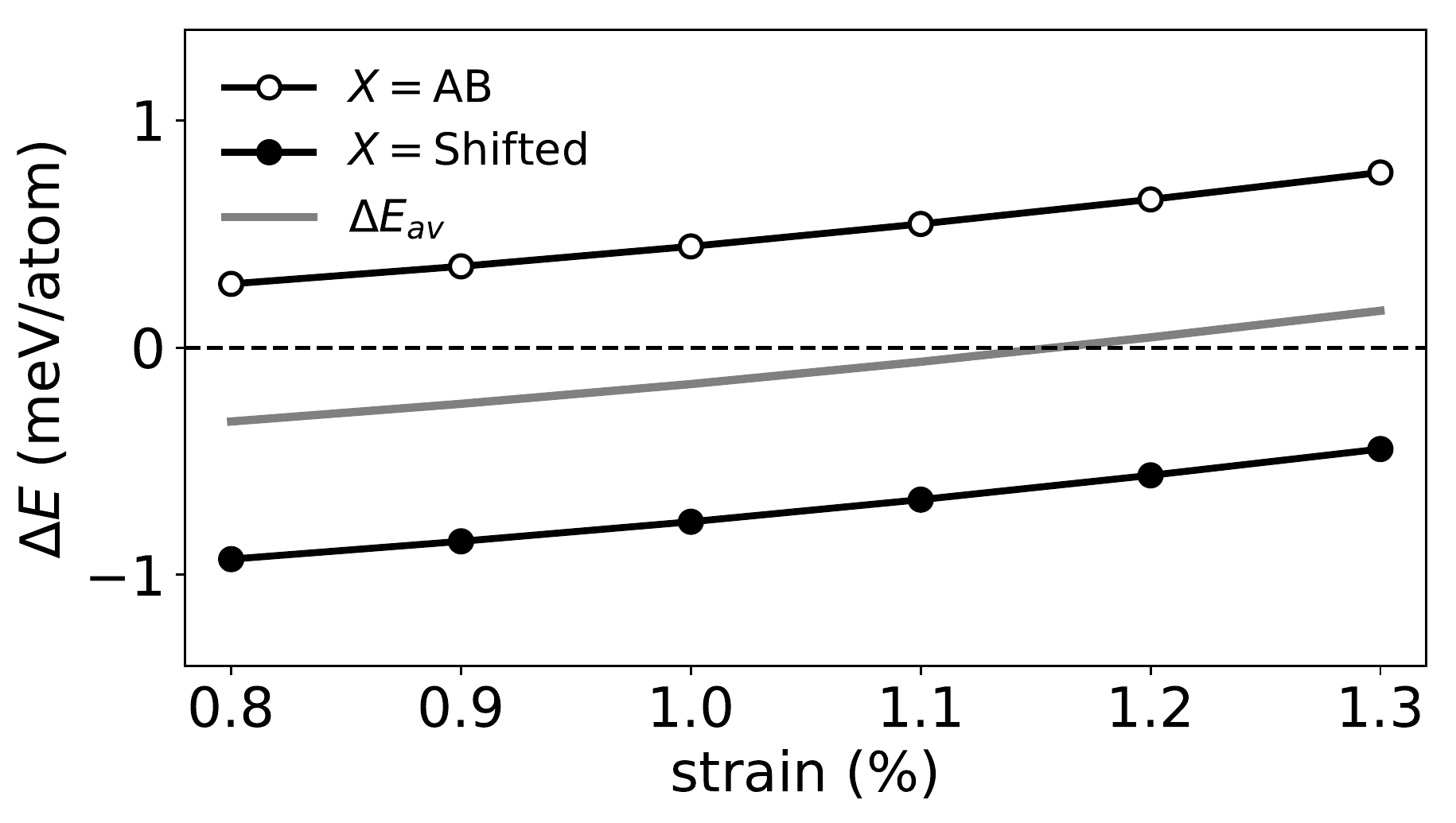}
\caption{Energy difference $\Delta E$ between the strained and unstrained free-layer geometries as a function of ZZ uniaxial strain applied to the bottom layer. In the strained case, the layers are AB-stacked, whereas in the unstrained case different stackings are possible, with different contributions to the $\Delta E$. These calculations were performed using the $6$-AGNR/MLG system. The averaged curve indicates a critical strain of $\sim1.2\%$.}
\label{fig:strain}
\end{figure}

The full heterostrained system is considered as an average of the two stacking extremes, AB and Shifted (more general cases will be discussed later): 
\begin{equation}
  \Delta E_\mathrm{av} = E_{\mathrm{AB,\,strain}} - \tfrac{1}{2}( E_{\mathrm{AB,\,unstrain}} + E_{\mathrm{shifted}\mathrm{,\,unstrain}}). 
  \label{eqn:deltaE1}
\end{equation}
Negative values of $\Delta E_\mathrm{av}$ imply that the free layer prefers to be strained so that the bilayer system remains AB-stacked. 
Positive values indicate that the free layer prefers to be unstrained, and the system adopts a non-uniform stacking profile. 
Figure \ref{fig:strain} shows $\Delta E$ and $\Delta E_\mathrm{av}$, as a function of strain applied along the ZZ direction.
The curve for the AB-stacked AGNR/MLG system shows that it prefers to be unstrained, i.e., that it is energetically favourable for the AGNR to break perfect AB-stacking, instead of maintaining a strain of between 0.8\% - 1.3\%.
This is not surprising, as the stacking mismatch is not too significant in this part of the modulated structure (c.f.~figure \ref{fig:blg}). 
However, in the shifted region, the stacking deviates furthest from AB when the free layer is unstrained, and the associated $\Delta E$ curve shows that a strained, uniform AB-stacking is preferred.
The overall preference of the system is a competition between these different regions.
For a uniform distribution of stackings in the modulated structure, the averaged case shown by the grey curve in Figure \ref{fig:strain} indicates that a transition occurs at a critical strain of $\sim 1.2\%$. This is in very good agreement with the estimate given by the simple model.

\section{Discussion}\label{sec:4}
Modulated stacking profiles in BLG systems lead to a wide range of new properties, particularly when combined with an interlayer bias.
Such a bias will open a band gap in regions with particular stackings, leading to a complex distribution of gapped and conducting regions which follow the underlying stacking pattern.
This has been widely studied in twisted bilayers where, for example, networks of 1D topological channels have been found between gapped AB- and BA- stacked regions of the Moir{\'e} pattern~\cite{san2013helical, PhysRevB.98.035404, PhysRevLett.121.037702}.
The localisation of electronic states in different regions of the lattice, together with the resulting strong interaction effects, are connected with the formation of correlated insulating states and unconventional superconductivity in twisted systems~\cite{stepanov2020untying}.

Heterostrained, untwisted BLG, as considered here, could potentially host a similar range of phenomena.
The schematic structure in figure \ref{fig:blg} shows the formation of a 1D Moir{\'e} pattern with different stacking profiles, which would also create a spatially-varying band gap landscape in the presence of an interlayer bias.
To maintain periodicity for our calculations, we neglected a Poisson contraction perpendicular to the applied strain.
Including such a contraction would lead to a 2D modulation of the stacking pattern and a more complete analogy with twisted systems.
We note that setting $\nu\neq0$ allows a wider range of stackings, when the free layer is unstrained, than for the simple 1D modulation considered in figures \ref{fig:blg} and \ref{fig:strain}.
In particular, more energetically unfavourable stackings, such as AA, are now possible, which may slightly increase the critical strain due to the increased energetic cost of breaking uniform stacking.
    
Finally, the estimate of the critical strain in this work is based on the free layer being either unstrained, or uniformly adopting the same strain as the bottom layer.
We have not considered, for example, an intermediate value of strain in the free layer.
We expect such cases to be less energetically favourable than the unstrained free layer.
This is because such cases will have to pay both strain and stacking-related energy costs, and the latter is expected to be largely strain independent, as discussed in Section \ref{sub:1}. 
Therefore the minimum energy heterostrained system should be that which minimises $\Delta E_\mathrm{strain}$, namely that with an unstrained free layer.
Due to the periodicity constraints of our calculations, we also not been able to explicitly consider the role of non-uniform strain in the free layer. Such non-uniform strain profiles could arise due to lattice relaxations, as occurs in TBLG~\cite{Gargiulo_2017, Carr2018:relaxationTBLG, kazmierczak2021strain}, where it serves to reduce the area of the AA-stacking region and maximize the area of the lower energy AB-stacking region~\cite{Nguyen2017}.

\section{Conclusions}\label{sec:5}

Our results strongly indicate the existence of a critical strain of $\sim 1\%$ which, when applied to one layer of BLG, can be used to tune the stacking profile of the system. 
Below the critical strain, it is energetically favourable to transfer the strain to the second layer in order to maintain a uniform AB-stacking configuration.
Above the critical value, the cost of maintaining the strain in the second layer is too high and the system prefers to release it and adopt a non-uniform stacking profile.
This finding is supported by DFT calculations which consider the energetic costs of strain and stacking independently in infinite, periodic systems, and by further calculations which consider both contributions simultaneously when one layer is represented by finite AGNRs.

\section*{Acknowledgements}
The authors are grateful for the financial support from Advanced Materials and Bioengineering Research centre (AMBER) (Grant No. caffreyn-AMBER 2-PF-Pillar-phD4) and Science Foundation Ireland (SFI). N.M.C.~was supported by a Science Foundation Ireland Starting Investigator Research Grant (15/SIRG/3314).
S.R.P. wishes to acknowledge funding from the Irish Research Council under the Laureate awards programme. 
Computational resources have been provided by Trinity Centre for High Performance Computing (TCHPC) and by Irish Centre for High-End Computing (ICHEC).

\addcontentsline{toc}{chapter}{References}
\newcommand{\newblock}{}
\bibliographystyle{unsrt}
\bibliography{References}

\begin{thebibliography}{10}

\bibitem{akinwande2017review}
Deji Akinwande, Christopher~J Brennan, J~Scott Bunch, Philip Egberts,
  Jonathan~R Felts, Huajian Gao, Rui Huang, Joon-Seok Kim, Teng Li, Yao Li,
  et~al.
\newblock A review on mechanics and mechanical properties of 2d
  materials—graphene and beyond.
\newblock {\em Extreme Mechanics Letters}, 13:42--77, 2017.

\bibitem{kim2020heterogeneous}
Jin~Myung Kim, Chullhee Cho, Ezekiel~Y Hsieh, and SungWoo Nam.
\newblock Heterogeneous deformation of two-dimensional materials for emerging
  functionalities.
\newblock {\em Journal of materials research}, 35(11):1369--1385, 2020.

\bibitem{novoselov2005two}
Kostya~S Novoselov, D~Jiang, F~Schedin, TJ~Booth, VV~Khotkevich, SV~Morozov,
  and Andre~K Geim.
\newblock Two-dimensional atomic crystals.
\newblock {\em Proceedings of the National Academy of Sciences},
  102(30):10451--10453, 2005.

\bibitem{katsnelson2007graphene}
Mikhail~I Katsnelson.
\newblock Graphene: carbon in two dimensions.
\newblock {\em Materials today}, 10(1-2):20--27, 2007.

\bibitem{lee2008measurement}
Changgu Lee, Xiaoding Wei, Jeffrey~W Kysar, and James Hone.
\newblock Measurement of the elastic properties and intrinsic strength of
  monolayer graphene.
\newblock {\em science}, 321(5887):385--388, 2008.

\bibitem{androulidakis2018tailoring}
Charalampos Androulidakis, Kaihao Zhang, Matthew Robertson, and Sameh Tawfick.
\newblock Tailoring the mechanical properties of 2d materials and
  heterostructures.
\newblock {\em 2D Materials}, 5(3):032005, 2018.

\bibitem{wang2014two}
Hong Wang, Fucai Liu, Wei Fu, Zheyu Fang, Wu~Zhou, and Zheng Liu.
\newblock Two-dimensional heterostructures: fabrication, characterization, and
  application.
\newblock {\em Nanoscale}, 6(21):12250--12272, 2014.

\bibitem{liu2014elastic}
Kai Liu, Qimin Yan, Michelle Chen, Wen Fan, Yinghui Sun, Joonki Suh, Deyi Fu,
  Sangwook Lee, Jian Zhou, Sefaattin Tongay, et~al.
\newblock Elastic properties of chemical-vapor-deposited monolayer mos2, ws2,
  and their bilayer heterostructures.
\newblock {\em Nano letters}, 14(9):5097--5103, 2014.

\bibitem{novoselov2004electric}
Kostya~S Novoselov, Andre~K Geim, Sergei~V Morozov, De-eng Jiang, Yanshui
  Zhang, Sergey~V Dubonos, Irina~V Grigorieva, and Alexandr~A Firsov.
\newblock Electric field effect in atomically thin carbon films.
\newblock {\em science}, 306(5696):666--669, 2004.

\bibitem{ohta2006controlling}
Taisuke Ohta, Aaron Bostwick, Thomas Seyller, Karsten Horn, and Eli Rotenberg.
\newblock Controlling the electronic structure of bilayer graphene.
\newblock {\em Science}, 313(5789):951--954, 2006.

\bibitem{mccann2013electronic}
Edward McCann and Mikito Koshino.
\newblock The electronic properties of bilayer graphene.
\newblock {\em Reports on Progress in physics}, 76(5):056503, 2013.

\bibitem{rozhkov2016electronic}
Alexandr~Vladimirovich Rozhkov, AO~Sboychakov, AL~Rakhmanov, and Franco Nori.
\newblock Electronic properties of graphene-based bilayer systems.
\newblock {\em Physics Reports}, 648:1--104, 2016.

\bibitem{ho2006coulomb}
Jon-Hsu Ho, CL~Lu, CC~Hwang, CP~Chang, and Min-Fa Lin.
\newblock Coulomb excitations in aa-and ab-stacked bilayer graphites.
\newblock {\em Physical Review B}, 74(8):085406, 2006.

\bibitem{castro2007biased}
Eduardo~V Castro, KS~Novoselov, SV~Morozov, NMR Peres, JMB~Lopes Dos~Santos,
  Johan Nilsson, F~Guinea, AK~Geim, and AH~Castro Neto.
\newblock Biased bilayer graphene: semiconductor with a gap tunable by the
  electric field effect.
\newblock {\em Physical review letters}, 99(21):216802, 2007.

\bibitem{silva2020electronic}
EL~Silva, MC~Santos, JM~Skelton, Tao Yang, T~Santos, SC~Parker, and A~Walsh.
\newblock Electronic and phonon instabilities in bilayer graphene under applied
  external bias.
\newblock {\em Materials Today: Proceedings}, 20:373--382, 2020.

\bibitem{bhattacharyya2016lifshitz}
Swastibrata Bhattacharyya and Abhishek~K Singh.
\newblock Lifshitz transition and modulation of electronic and transport
  properties of bilayer graphene by sliding and applied normal compressive
  strain.
\newblock {\em Carbon}, 99:432--438, 2016.

\bibitem{son2011electronic}
Young-Woo Son, Seon-Myeong Choi, Yoon~Pyo Hong, Sungjong Woo, and Seung-Hoon
  Jhi.
\newblock Electronic topological transition in sliding bilayer graphene.
\newblock {\em Physical Review B}, 84(15):155410, 2011.

\bibitem{dos2007graphene}
JMB~Lopes Dos~Santos, NMR Peres, and AH~Castro Neto.
\newblock Graphene bilayer with a twist: Electronic structure.
\newblock {\em Physical review letters}, 99(25):256802, 2007.

\bibitem{shallcross2010electronic}
S~Shallcross, S~Sharma, E~Kandelaki, and OA~Pankratov.
\newblock Electronic structure of turbostratic graphene.
\newblock {\em Physical Review B}, 81(16):165105, 2010.

\bibitem{dai2016twisted}
Shuyang Dai, Yang Xiang, and David~J Srolovitz.
\newblock Twisted bilayer graphene: Moir{\'e} with a twist.
\newblock {\em Nano letters}, 16(9):5923--5927, 2016.

\bibitem{mcgilly2020visualization}
Leo~J McGilly, Alexander Kerelsky, Nathan~R Finney, Konstantin Shapovalov,
  En-Min Shih, Augusto Ghiotto, Yihang Zeng, Samuel~L Moore, Wenjing Wu, Yusong
  Bai, et~al.
\newblock Visualization of moir{\'e} superlattices.
\newblock {\em Nature Nanotechnology}, 15(7):580--584, 2020.

\bibitem{ni2008uniaxial}
Zhen~Hua Ni, Ting Yu, Yun~Hao Lu, Ying~Ying Wang, Yuan~Ping Feng, and Ze~Xiang
  Shen.
\newblock Uniaxial strain on graphene: Raman spectroscopy study and band-gap
  opening.
\newblock {\em ACS nano}, 2(11):2301--2305, 2008.

\bibitem{pereira2009tight}
Vitor~M Pereira, AH~Castro Neto, and NMR Peres.
\newblock Tight-binding approach to uniaxial strain in graphene.
\newblock {\em Physical Review B}, 80(4):045401, 2009.

\bibitem{ni2010anisotropic}
Zhonghua Ni, Hao Bu, Min Zou, Hong Yi, Kedong Bi, and Yunfei Chen.
\newblock Anisotropic mechanical properties of graphene sheets from molecular
  dynamics.
\newblock {\em Physica B: Condensed Matter}, 405(5):1301--1306, 2010.

\bibitem{pellegrino2010strain}
FMD Pellegrino, GGN Angilella, and R~Pucci.
\newblock Strain effect on the optical conductivity of graphene.
\newblock {\em Physical Review B}, 81(3):035411, 2010.

\bibitem{mucha2011strained}
Marcin Mucha-Kruczy{\'n}ski, Igor~L Aleiner, and Vladimir~I Fal’ko.
\newblock Strained bilayer graphene: Band structure topology and landau level
  spectrum.
\newblock {\em Physical Review B}, 84(4):041404, 2011.

\bibitem{gradinar2012conductance}
Diana~A Gradinar, Henning Schomerus, and Vladimir~I Fal'Ko.
\newblock Conductance anomaly near the lifshitz transition in strained bilayer
  graphene.
\newblock {\em Physical Review B}, 85(16):165429, 2012.

\bibitem{de2012space}
Fernando de~Juan, Mauricio Sturla, and Maria~AH Vozmediano.
\newblock Space dependent fermi velocity in strained graphene.
\newblock {\em Physical review letters}, 108(22):227205, 2012.

\bibitem{zhang2014transport}
Hang Zhang, Jhao-Wun Huang, Jairo Velasco~Jr, Kevin Myhro, Matt Maldonado,
  David~Dung Tran, Zeng Zhao, Fenglin Wang, Yongjin Lee, Gang Liu, et~al.
\newblock Transport in suspended monolayer and bilayer graphene under strain: a
  new platform for material studies.
\newblock {\em Carbon}, 69:336--341, 2014.

\bibitem{Settnes2016}
Mikkel Settnes, Stephen~R. Power, Mads Brandbyge, and Antti-Pekka Jauho.
\newblock Graphene nanobubbles as valley filters and beam splitters.
\newblock {\em Phys. Rev. Lett.}, 117:276801, Dec 2016.

\bibitem{aitken2010effects}
Zachary~H Aitken and Rui Huang.
\newblock Effects of mismatch strain and substrate surface corrugation on
  morphology of supported monolayer graphene.
\newblock {\em Journal of Applied Physics}, 107(12):123531, 2010.

\bibitem{yang2021strain}
Shengxue Yang, Yujia Chen, and Chengbao Jiang.
\newblock Strain engineering of two-dimensional materials: Methods, properties,
  and applications.
\newblock {\em InfoMat}, 3(4):397--420, 2021.

\bibitem{yu2008raman}
Ting Yu, Zhenhua Ni, Chaoling Du, Yumeng You, Yingying Wang, and Zexiang Shen.
\newblock Raman mapping investigation of graphene on transparent flexible
  substrate: the strain effect.
\newblock {\em The Journal of Physical Chemistry C}, 112(33):12602--12605,
  2008.

\bibitem{huang2009phonon}
Mingyuan Huang, Hugen Yan, Changyao Chen, Daohua Song, Tony~F Heinz, and James
  Hone.
\newblock Phonon softening and crystallographic orientation of strained
  graphene studied by raman spectroscopy.
\newblock {\em Proceedings of the National Academy of Sciences},
  106(18):7304--7308, 2009.

\bibitem{androulidakis2020tunable}
Charalampos Androulidakis, Emmanuel~N Koukaras, George Paterakis, George
  Trakakis, and Costas Galiotis.
\newblock Tunable macroscale structural superlubricity in two-layer graphene
  via strain engineering.
\newblock {\em Nature communications}, 11(1):1--11, 2020.

\bibitem{frank2012phonon}
Otakar Frank, Milan Bousa, Ibtsam Riaz, Rashid Jalil, Kostya~S Novoselov,
  Georgia Tsoukleri, John Parthenios, Ladislav Kavan, Konstantinos Papagelis,
  and Costas Galiotis.
\newblock Phonon and structural changes in deformed bernal stacked bilayer
  graphene.
\newblock {\em Nano Letters}, 12(2):687--693, 2012.

\bibitem{wang2019robust}
Kunqi Wang, Wengen Ouyang, Wei Cao, Ming Ma, and Quanshui Zheng.
\newblock Robust superlubricity by strain engineering.
\newblock {\em Nanoscale}, 11(5):2186--2193, 2019.

\bibitem{choi2010controlling}
Seon-Myeong Choi, Seung-Hoon Jhi, and Young-Woo Son.
\newblock Controlling energy gap of bilayer graphene by strain.
\newblock {\em Nano letters}, 10(9):3486--3489, 2010.

\bibitem{van2016piezoelectricity}
Matthias Van~der Donck, C~De~Beule, B~Partoens, FM~Peeters, and B~Van~Duppen.
\newblock Piezoelectricity in asymmetrically strained bilayer graphene.
\newblock {\em 2D Materials}, 3(3):035015, 2016.

\bibitem{parhizkar2022strained}
Alireza Parhizkar and Victor Galitski.
\newblock Strained bilayer graphene, emergent energy scales, and moire gravity.
\newblock {\em Physical Review Research}, 4(2):L022027, 2022.

\bibitem{crosse2014strain}
JA~Crosse.
\newblock Strain-dependent conductivity in biased bilayer graphene.
\newblock {\em Physical Review B}, 90(23):235403, 2014.

\bibitem{bistritzer2011moire}
Rafi Bistritzer and Allan~H MacDonald.
\newblock Moir{\'e} bands in twisted double-layer graphene.
\newblock {\em Proceedings of the National Academy of Sciences},
  108(30):12233--12237, 2011.

\bibitem{kresse1993ab}
Georg Kresse and J{\"u}rgen Hafner.
\newblock Ab initio molecular dynamics for liquid metals.
\newblock {\em Physical review B}, 47(1):558, 1993.

\bibitem{kresse1994ab}
Georg Kresse and J{\"u}rgen Hafner.
\newblock Ab initio molecular-dynamics simulation of the
  liquid-metal--amorphous-semiconductor transition in germanium.
\newblock {\em Physical Review B}, 49(20):14251, 1994.

\bibitem{kresse1996efficiency}
Georg Kresse and J{\"u}rgen Furthm{\"u}ller.
\newblock Efficiency of ab-initio total energy calculations for metals and
  semiconductors using a plane-wave basis set.
\newblock {\em Computational materials science}, 6(1):15--50, 1996.

\bibitem{kresse1996efficient}
Georg Kresse and J{\"u}rgen Furthm{\"u}ller.
\newblock Efficient iterative schemes for ab initio total-energy calculations
  using a plane-wave basis set.
\newblock {\em Physical review B}, 54(16):11169, 1996.

\bibitem{blochl1994projector}
Peter~E Bl{\"o}chl.
\newblock Projector augmented-wave method.
\newblock {\em Physical review B}, 50(24):17953, 1994.

\bibitem{kresse1999ultrasoft}
Georg Kresse and Daniel Joubert.
\newblock From ultrasoft pseudopotentials to the projector augmented-wave
  method.
\newblock {\em Physical review b}, 59(3):1758, 1999.

\bibitem{perdew1996generalized}
John~P Perdew, Kieron Burke, and Matthias Ernzerhof.
\newblock Generalized gradient approximation made simple.
\newblock {\em Physical review letters}, 77(18):3865, 1996.

\bibitem{grimme2006semiempirical}
Stefan Grimme.
\newblock Semiempirical gga-type density functional constructed with a
  long-range dispersion correction.
\newblock {\em Journal of computational chemistry}, 27(15):1787--1799, 2006.

\bibitem{tao2012comparative}
Wang Tao, Guo Qing, Liu Yan, and Sheng Kuang.
\newblock A comparative investigation of an ab-and aa-stacked bilayer graphene
  sheet under an applied electric field: a density functional theory study.
\newblock {\em Chinese Physics B}, 21(6):067301, 2012.

\bibitem{lee2008growth}
Jae-Kap Lee, Seung-Cheol Lee, Jae-Pyoung Ahn, Soo-Chul Kim, John~IB Wilson, and
  Phillip John.
\newblock The growth of aa graphite on (111) diamond.
\newblock {\em The Journal of chemical physics}, 129(23):234709, 2008.

\bibitem{dahn1990suppression}
JR~Dahn, Rosamaria Fong, and MJ~Spoon.
\newblock Suppression of staging in lithium-intercalated carbon by disorder in
  the host.
\newblock {\em Physical Review B}, 42(10):6424, 1990.

\bibitem{raza2011edge}
Hassan Raza.
\newblock Edge and passivation effects in armchair graphene nanoribbons.
\newblock {\em Physical Review B}, 84(16):165425, 2011.

\bibitem{nakada1996edge}
Kyoko Nakada, Mitsutaka Fujita, Gene Dresselhaus, and Mildred~S Dresselhaus.
\newblock Edge state in graphene ribbons: Nanometer size effect and edge shape
  dependence.
\newblock {\em Physical Review B}, 54(24):17954, 1996.

\bibitem{fujita1996peculiar}
Mitsutaka Fujita, Katsunori Wakabayashi, Kyoko Nakada, and Koichi Kusakabe.
\newblock Peculiar localized state at zigzag graphite edge.
\newblock {\em Journal of the Physical Society of Japan}, 65(7):1920--1923,
  1996.

\bibitem{wakabayashi1999electronic}
Katsunori Wakabayashi, Mitsutaka Fujita, Hiroshi Ajiki, and Manfred Sigrist.
\newblock Electronic and magnetic properties of nanographite ribbons.
\newblock {\em Physical Review B}, 59(12):8271, 1999.

\bibitem{san2013helical}
Pablo San-Jose and Elsa Prada.
\newblock Helical networks in twisted bilayer graphene under interlayer bias.
\newblock {\em Physical Review B}, 88(12):121408, 2013.

\bibitem{PhysRevB.98.035404}
Dmitry~K. Efimkin and Allan~H. MacDonald.
\newblock Helical network model for twisted bilayer graphene.
\newblock {\em Phys. Rev. B}, 98:035404, 2018.

\bibitem{PhysRevLett.121.037702}
Shengqiang Huang, Kyounghwan Kim, Dmitry~K. Efimkin, Timothy Lovorn, Takashi
  Taniguchi, Kenji Watanabe, Allan~H. MacDonald, Emanuel Tutuc, and Brian~J.
  LeRoy.
\newblock Topologically protected helical states in minimally twisted bilayer
  graphene.
\newblock {\em Phys. Rev. Lett.}, 121:037702, 2018.

\bibitem{stepanov2020untying}
Petr Stepanov, Ipsita Das, Xiaobo Lu, Ali Fahimniya, Kenji Watanabe, Takashi
  Taniguchi, Frank~HL Koppens, Johannes Lischner, Leonid Levitov, and Dmitri~K
  Efetov.
\newblock Untying the insulating and superconducting orders in magic-angle
  graphene.
\newblock {\em Nature}, 583(7816):375--378, 2020.

\bibitem{Gargiulo_2017}
Fernando Gargiulo and Oleg~V Yazyev.
\newblock Structural and electronic transformation in low-angle twisted bilayer
  graphene.
\newblock {\em 2D Materials}, 5(1):015019, 2017.

\bibitem{Carr2018:relaxationTBLG}
Stephen Carr, Daniel Massatt, Steven~B. Torrisi, Paul Cazeaux, Mitchell Luskin,
  and Efthimios Kaxiras.
\newblock Relaxation and domain formation in incommensurate two-dimensional
  heterostructures.
\newblock {\em Phys. Rev. B}, 98:224102, 2018.

\bibitem{kazmierczak2021strain}
Nathanael~P Kazmierczak, Madeline Van~Winkle, Colin Ophus, Karen~C Bustillo,
  Stephen Carr, Hamish~G Brown, Jim Ciston, Takashi Taniguchi, Kenji Watanabe,
  and D~Kwabena Bediako.
\newblock Strain fields in twisted bilayer graphene.
\newblock {\em Nature materials}, 20(7):956--963, 2021.

\bibitem{Nguyen2017}
Nguyen N.~T. Nam and Mikito Koshino.
\newblock Lattice relaxation and energy band modulation in twisted bilayer
  graphene.
\newblock {\em Phys. Rev. B}, 96(12):075311, 2017.

\end{thebibliography}


\end{document}